\documentclass[sigconf]{acmart}

\setcopyright{none}

\copyrightyear{2023} 
\acmYear{2023} 
\setcopyright{acmlicensed}\acmConference[CHIIR '23]{ACM SIGIR Conference on Human Information Interaction and Retrieval}{March 19--23, 2023}{Austin, TX, USA}
\acmBooktitle{ACM SIGIR Conference on Human Information Interaction and Retrieval (CHIIR '23), March 19--23, 2023, Austin, TX, USA}
\acmPrice{15.00}
\acmDOI{10.1145/3576840.3578290}
\acmISBN{979-8-4007-0035-4/23/03}

\begin{document}

\title[Cognitive load in fact-checking]{True or false? Cognitive load when reading COVID-19 news headlines: an eye-tracking study}

\author{Li Shi}
\orcid{0000-0002-8252-4363}
\affiliation{%
  \department{IX Lab, School of Information}
  \institution{The University of Texas at Austin}
  \city{Austin}
  \state{Texas}
  \country{USA}
}
\email{lilylashi@utexas.edu}

\author{Nilavra Bhattacharya}
\orcid{0000-0001-7864-7726}
\affiliation{%
  \department{IX Lab, School of Information}
  \institution{ The University of Texas at Austin}
  \city{Austin}
  \state{Texas}
  \country{USA}}
\email{nilavra@ieee.org}

\author{Anubrata Das}
\orcid{0000-0002-5412-6149}
\affiliation{%
  \department{School of Information}
  \institution{The University of Texas at Austin}
  \city{Austin}
  \state{Texas}
  \country{USA}}
\email{anubrata@utexas.edu}

\author{Jacek Gwizdka}
\orcid{0000-0003-2273-3996}
\affiliation{%
  \department{IX Lab, School of Information}
  \institution{The University of Texas at Austin}
  \city{Austin}
  \state{Texas}
  \country{USA}}
\email{jacekg@utexas.edu}

\renewcommand{\shortauthors}{Shi, et al.}

\begin{abstract}
Misinformation is an important topic in the Information Retrieval (IR) context and has implications for both system-centered and user-centered IR. While it has been established that the performance in discerning misinformation is affected by a person's cognitive load, the variation in cognitive load in judging the veracity of news is less understood.  To understand the variation in cognitive load imposed by reading news headlines related to COVID-19 claims, within the context of a fact-checking system, we conducted a within-subject, lab-based, quasi-experiment (N=40) with eye-tracking. Our results suggest that examining true claims imposed a higher cognitive load on participants when news headlines provided incorrect evidence for a claim and were inconsistent with the person's prior beliefs. In contrast, checking false claims imposed a higher cognitive load when the news headlines provided correct evidence for a claim and were consistent with the participants' prior beliefs. However, changing beliefs after examining a claim did not have a significant relationship with cognitive load while reading the news headlines. The results illustrate that reading news headlines related to true and false claims in the fact-checking context impose different levels of cognitive load.
Our findings suggest that user engagement with tools for discerning misinformation needs to account for the possible variation in the mental effort involved in different information contexts. 

% Our findings suggest that to encourage people to discern misinformation, measures should be taken to calibrate the cognitive load according to different information contexts.

% <FOR_LS> Cognitive load plays a role in a person's ability to discern misinformation, but the variation in cognitive load while evaluating the veracity of news is not well understood. To address this gap in knowledge, we conducted a lab-based quasi-experiment (N=40) using eye-tracking to observe cognitive load while reading COVID-19-related news headlines within the context of a fact-checking system. We found that examining true claims imposed higher cognitive load when the headlines provided incorrect evidence and were inconsistent with the participant's prior beliefs, while checking false claims imposed higher cognitive load when the headlines provided correct evidence and were consistent with the participant's prior beliefs. However, changing beliefs after examining a claim did not significantly affect cognitive load while reading the headlines. These results suggest that different levels of cognitive load are imposed when reading true and false claims in the context of a fact-checking system, and that the cognitive load should be carefully calibrated based on the specific information context to encourage people to discern misinformation.
\end{abstract}

\begin{CCSXML}
<ccs2012>
   <concept>
       <concept_id>10002951.10003317.10003331.10003336</concept_id>
       <concept_desc>Information systems~Search interfaces</concept_desc>
       <concept_significance>500</concept_significance>
       </concept>
   <concept>
       <concept_id>10003120.10003121.10003122.10003334</concept_id>
       <concept_desc>Human-centered computing~User studies</concept_desc>
       <concept_significance>500</concept_significance>
       </concept>
   <concept>
       <concept_id>10003120.10003123.10011759</concept_id>
       <concept_desc>Human-centered computing~Empirical studies in interaction design</concept_desc>
       <concept_significance>500</concept_significance>
       </concept>
 </ccs2012>
\end{CCSXML}

\ccsdesc[500]{Information systems~Search interfaces}
\ccsdesc[500]{Human-centered computing~User studies}
\ccsdesc[500]{Human-centered computing~Empirical studies in interaction design}

\keywords{fact checking, misinformation, cognitive load, pupil dilation}


% % ----- fig:interface -----
%   \centering
%   \includegraphics[width=\linewidth]{figs/interface.pdf}
%   \caption{Interface of the mock fact-checking system used in the study. 
%   Red frames represent the areas of interest (AOIs) around the news headlines, that are used in the analysis.}
% %   \Description{}
%   \label{fig:interface}
% % ----- fig:interface -----


\maketitle
% -----------------------------------
\section{Introduction}
\label{sec:intro}
With increasing access to information technologies, misinformation becomes more easily and widely disseminated through social media \cite{del2016spreading}. Research shows that misinformation spreads more rapidly than true information. Widespread misinformation can potentially disrupt public health, democratic processes, and public discourse \cite{xie2020global,hochschild2015facts}. Consequently, there is a growing public and academic interest in tackling the challenges associated with misinformation. Specifically, fact-checking as a method for curbing misinformation has gained attention in research and practice \cite{graves2019fact,Arnold2020}. 

While we have seen a surge of fact-checking organizations in the past few years, addressing the gigantic scale of misinformation online may be intractable with expert fact-checkers alone. Consequently, we have seen interest in automating the fact-checking process, either towards building end-to-end automation, or as decision support systems for fact-checkers \cite{Nakov2021AutomatedFF,Guo2021ASO,graves2018understanding}. 

Automated fact-checking research are often evaluated using automated metrics. However, the primary goal of automated fact-checking tools are to help fact-checkers with their decision making. Recent research has started looking into several human factors associated with fact-checking such as usability, trust, and intelligibility \cite{DAS2023103219,nguyen2018believe,Shi2022effect,jiang2020factoring}. Additionally, some research has also analyzed the factors affecting human judgment during the interaction with misinformation, including the reading behaviors and beliefs in the misinformation \cite{ecker2022psychological, roozenbeek2020susceptibility}. However, a detailed understanding of users cognitive load and fact-checking system have not been studied in great detail.

Previous research in cognition and discerning misinformation typically displayed one news headline per trial. However, in realistic information search scenarios, people often encounter several pieces of information in one search. Moreover, the headlines shown in the previous experiments were mostly related to political topics. Currently, misinformation not only threatens democracy but also public health regarding the situation of the COVID-19 pandemic. As stated in \cite{zarocostas2020fight}, ``we're not just fighting an epidemic; we're fighting an infodemic''. Therefore, we designed a lab-based experiment where users checked COVID-19-related health claims in a fact-checking system and were shown multiple news headlines related to the claim simultaneously in one screen. During the experimental session, an eye-tracker recorded their pupillary response.
The aim of the study was to examine how the cognitive load is impacted when users read the news headlines in a fact-checking context, and how it is related to their belief change and misinformation judgment.
The contributions of this study include: 
(1) examining the effectiveness of pupil dilation measurements as an indicator of cognitive load in reading news headlines;
(2) comparing the cognitive load imposed by reading news headlines under varied conditions of claim correctness, headline-belief stance, and evidence correctness;
and
(3) developing the understanding of the cognitive processes in discerning misinformation by investigating the cognitive load in a more realistic scenario.

% misinformation draws more and more attention in information search and social media
% previous research starting looking into the cognitive load in reading and discerning fake news

% [Previous research mostly focuses on political misinformation. Our research will look into another important area - health information]
% [Previous research only displays one news headline per time. Our research will study the performance in a more realistic context. Several news headlines at one time.]

% -----------------------------------
\section{Background}
\label{sec:bg}

\subsection{Human Evaluation of Automated Fact-Checking}

Several studies have evaluated automated fact-checking from a human factor perspective. Such factors include understanding, usability, intelligibility, and trust in those systems \cite{nguyen2018believe,Shi2022effect,mohseni2021machine,das2019cobweb}. \citet{nguyen2018believe} studied the effect of intelligibility in automated fact-checking on user trust and task accuracy. \citet{mohseni2021machine} examined the effect of intelligibility in calibrating user trust on automated fact-checking systems. Complementary to that, \citet{das2019cobweb} investigated the role of confirmation bias in using automated fact-checking systems. The studies mentioned above focused on measures related to the fact-checking task and not on the user behavior while interacting with the system.

In contrast, \citet{Shi2022effect} examined user interaction with the fact-checking interface. They studied the effect of interactivity on several factors associated with user interaction, e.g., dwell time, attention, and mental resources, with the help of eye tracking. Our work extends such research and employs similar eye-tracking methodology in investigating users' cognitive load while interacting with automated fact-checking systems..  

\subsection{Cognitive load and discerning misinformation}
\label{sec:cog_misinfo}
\citeN{kahneman2011thinking}'s dual-process theory states that human cognition can be conceptualized as two systems, System 1 and System 2. 
System 1 ``operates automatically and quickly, with little or no effort and no sense of voluntary control'', i.e., these are autonomous and intuitive processes.
System 2 ``allocates attention to the effortful mental activities that demand it, including complex computations. The operations of System 2 are often associated with the subjective experience of agency, choice, and concentration'', i.e., these are deliberative and analytic processes. 
Previous research found that people who engage analytical thinking perform better on rational thinking tests \cite{stanovich2011complexity}. Similarly, the heuristic-systematic model explains that System 1 uses heuristics, while System 2 uses analysis, which makes people using System 1 more susceptible to decision-making biases \cite{chen1999motivated}. People who are engaging System 2 put conscious effort into thinking and think analytically, and thus are more likely to identify misinformation. To investigate it, \cite{pennycook2019lazy} used Cognitive Reflection Test (CRT) as a measure of the willingness of activating System 2 and found that CRT is positively correlated with the capability to discern fake news. Another study investigated the impact of deliberation on correcting intuitive mistakes \cite{bago2020fake}. Researchers found that when people have more time to reconsider false claims, they are less likely to trust them \cite{moravec2020appealing}.

Based on the dual-process theory \cite{kahneman2011thinking}, cognitive load can be used as an indication of System 2 activation, and thereby to study the performance and cognitive processes engaged in identifying fake news articles. 
\citeN{Mirhoseini_Early_Hassanein_2022} found that higher cognitive load was imposed when users have better performance in discerning misinformation.
Users with higher cognitive load utilize more System 2 resources, and deliberate and rationally examine the information correctness and ultimately discern misinformation. Additionally, pupillary response was shown to be a reliable physiological measure of cognitive load \cite{hossain2014understanding} since pupil dilation is associated with the amount of load on memory \cite{kahneman1966pupil}.
% A study on the fake news phenomenon using pupil metrics has found that people who perform better in identifying misinformation have higher cognitive load \cite{Mirhoseini_Early_Hassanein_2022}.
% Moreover, researchers measured cognitive load as an indication of System 2 activation, which enabled them to study the performance in identifying fake news articles. 

\subsection{Cognitive load and pupil dilation}
\label{sec:cog_pupil}

Cognitive load refers to the amount of working memory resources required to perform a cognitive task \cite{paas2016cognitive}. 
Typically there are three types of cognitive load measurements: task performance, subjective, and physiological \cite{o1986workload,gwizdka2021overloading}.
Task performance measures capture how well the user is performing a given task, such as task completion time, and the number of errors.
Subjective measures use self-rating scales of cognitive load, such as NASA-TLX questionnaire \cite{hart2006nasa}. These measures are simple to collect but cannot reflect rapid and dynamic cognitive load changes \cite{palinko2010estimating}.
Physiological measures include heart-rate variability (HRV), galvanic skin response (GSR), Electroencephalography (EEG), and eye-tracking measurements \cite{urrestilla2020measuring, antonenko2010using, shi2007galvanic}. Multiple eye-tracking measures, such as blink frequency and pupil dilation, have been shown to correlate with cognitive load levels \cite{siegle2008blink}.
In the past decades, researchers have found that the behavior of the pupil is a direct reflection of neurological and mental activity \cite{hess1964pupil}. \cite{kahneman1966pupil} showed that the changes of pupil diameter are related to task difficulties, and pupil dilation is associated with the amount of load on memory. Therefore, the pupillary response could be utilized as a reliable physiological measurement of cognitive load \cite{hossain2014understanding} in the misinformation studies.  

Various metrics are used to process pupil diameter data, and therefore to estimate mental workload. 
A common approach is to measure pupil dilation relative to a baseline. 
The baseline could be the average pupil diameter measured during a baseline trial \cite{kruger2013measuring} or during a baseline measurement made at the beginning of each trial \cite{krejtz2018eye}, or during whole experimental session of each participant \cite{gwizdka2014characterizing}.
Then the pupil size difference which is calculated with respect to the baseline is so called Relative Pupil Dilation (RPD) \cite{gwizdka2017temporal,wang2021pupillary}.
Another kind of alternative metric is proposed based on the moment-to-moment change in pupil diameter. 
This method estimates the frequency of pupil oscillation and fluctuation of pupil dilation while separating the effects of illumination. 
It was first proposed by Marshall with the measurement called the Index of Cognitive Activity (ICA) \cite{marshall2002index}. 
Since ICA is closed source, \citeN{duchowski2018index} offered a similar and open-source, fully-detailed measurement called the Index of Pupil Dilation (IPA). 
The researchers further proposed the Index of High/Low Pupillary Activity (LHIPA), by taking the tonic component (LF) into account, in addition to pupil phasic response (HF), which suggested to be a more reliable indicator of cognitive load \cite{duchowski2020low}. In this paper, we attempted to use the LHIPA and RPD as the indicators of cognitive load.

% <FOR_LS> Pupil diameter data is often analyzed using various metrics to estimate mental workload. One common method is to compare pupil dilation to a baseline, which can be the average pupil diameter measured during a baseline trial \cite{kruger2013measuring}, at the beginning of each trial \cite{krejtz2018eye}, or throughout the entire experimental session for each participant \cite{gwizdka2014characterizing}. The resulting difference in pupil size is known as the Relative Pupil Dilation (RPD) \cite{gwizdka2017temporal,wang2021pupillary}. Another approach involves analyzing the moment-to-moment changes in pupil diameter to estimate the frequency of pupil oscillation and fluctuation while accounting for the effects of illumination. This method, called the Index of Cognitive Activity (ICA) \cite{marshall2002index}, was proposed by Marshall. However, as ICA is a closed source, \citeN{duchowski2018index} developed a similar, open-source measure called the Index of Pupil Dilation (IPA). Additionally, the researchers proposed the Index of High/Low Pupillary Activity (LHIPA), which includes the tonic component (LF) in addition to pupil phasic response (HF) and is thought to be a more reliable indicator of cognitive load \cite{duchowski2020low}. In this study, we attempted to use both the LHIPA and RPD as indicators of cognitive load.

In previous research, eye-tracking was employed to investigate the effect of misinformation on cognitive activities. 
It was found that people fixated more frequently, had longer fixation duration, and increased pupil diameter when reading fake news compared to real news \cite{sumer2021fakenewsperception, hansen2020factuality, ladeira2022visual}. This is because reading false news imposed higher cognitive load on account of the reduced heuristic availability \cite{ladeira2022visual}.
Furthermore, researchers measured pupil dilation in investigating the performance of judging the accuracy of the headlines and demonstrated that pupils dilate more when people perform better on the misinformation judgment task \cite{Mirhoseini_Early_Hassanein_2022}. This study showed that higher cognitive load was associated with identifying misinformation.

% research question and hypothesis
Therefore, in our research, we measured pupil dilation as indication of cognitive load in information processing. We extended previous works to more realistic search scenarios, in which users encountered several relevant news headlines related to a single claim, identify misinformation, and determine the correctness of the claim. We aimed to explore how cognitive load is impacted in the fact-checking context (i.e., by the evidence correctness and users' prior beliefs), and if it was related to users' belief change. We hypothesized that:

\textbf{H1:} Reading news headlines that provide incorrect evidence imposes higher cognitive load.

\textbf{H2:} Reading news headlines that are inconsistent with their prior beliefs imposes higher cognitive load.

\textbf{H3:} Changing one's beliefs, and especially correcting beliefs, imposes higher cognitive load.
% Prior beliefs would influence the cognitive attention to the headlines. headlines that challenge their opinion receive little cognitive attention. \cite{moravec2018fake}

% ----- fig:interface -----
\begin{figure*}[h]
  \centering
  \includegraphics[width=\linewidth]{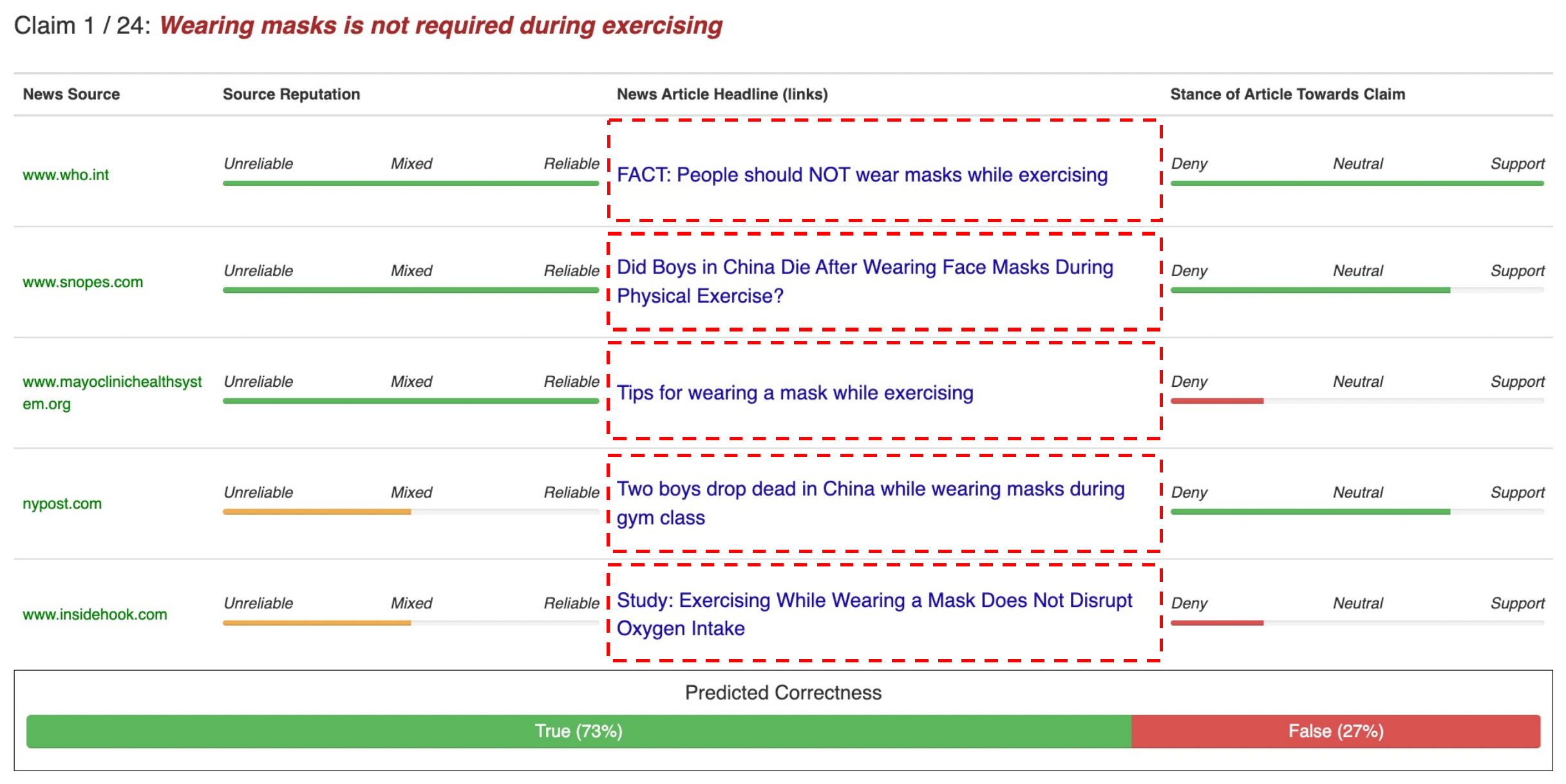}
  \caption{Interface of the mock fact-checking system used in the study. 
  Red frames represent the areas of interest (AOIs) around the news headlines, that are used in the analysis.}
%   \Description{}
  \label{fig:interface}
\end{figure*}
% ----- fig:interface -----

% -----------------------------------
\section{Methods}
\label{sec:methods}

\subsection{Experimental Design}
\label{sec:exp_design}
%% pre/post trial questionnaire, true/false/unsure claim (only select extreme instances)
A controlled, within-subjects eye-tracking study was conducted 
in a usability lab at a university, with
%in the Information eXperience usability lab at the University of Texas at Austin 
$N=40$ participants (22 females).
Participants interacted with a mock fact-checking system containing claims and news-article headlines in English language (Figure ~\ref{fig:interface}). 
Participants were pre-screened for native-level English familiarity, 20/20 vision (uncorrected or corrected), and non-expert topic familiarity of the content being shown in the fact checking system. 
Upon completion of the study, each participant was compensated with USD 25.

\subsection{Apparatus}
\label{sec:exp_appatratus}
A Tobii TX-300 eye-tracker was used to record participants' eye movements and pupil dilation. % iMotions 
Commercial usability and eye-tracking software was used to conduct the study, record raw gaze data, and perform partial data-cleaning and filtration for downstream analyses.
Data analysis was performed in Python and R languages.

\subsection{Mock Fact Checking System}
\label{sec:tasks_and_data_set}

Participants interacted with a mock fact checking system (Figure \ref{fig:interface}), and examined 24 COVID-19 related claims in the system.
% trials in each experiment session.
% In each interface, there were 12 trials. 
% from the two versions of the system 
Each claim was shown at the top of the interface. 
Surrogates of five related news articles were presented below the claim, each with its corresponding news source, source reputation, news headline, and the article’s stance towards the claim. 
Based on the article’s stance and news source reputation, the system provided a prediction of the claim’s correctness at the bottom. 
The news headlines were clickable, and upon clicking, opened the news article in a new browser tab.
Each claim examination consisted of viewing the claim, the headlines of the news articles, and, optionally, clicking the news articles to read them in detail. 
To mitigate the effect of background luminance of pupil dilation, the color and luminance of the fact-checking-system interface was kept constant during the experimental session.

The claims and corresponding news-articles were on the topic of the COVID-19 pandemic. 
They were handpicked by the researchers to simulate a COVID-19 fact-checking system for usability analysis. 
Each claim was selected so as to have a pre-assigned ground-truth correctness value of TRUE, FALSE, or UNSURE (for claims that are partially true, or not totally proven at the time of data collection).
The TRUE and UNSURE claims were handpicked from reputed websites in the medical domain, such as World Health Organization, WebMD, Mayoclinic, Johns Hopkins University, US State Government webpages, and others. 
The FALSE claims were sourced by searching for ``coronavirus myths'' on popular search engines. 
The relevant news articles for each claim were collected by manually searching the web.
The source reputations for news articles were collected from existing datasets \cite{gruppi_nela-gt-2019_2020,norregaard2019nela}, while the stance values of each news article towards each claim were labelled by the researchers. 
Two example claims are ``wearing masks is not required during exercising'', and ``asymptomatic people can transmit COVID-19''. 
In total there were 24 claims (8 TRUE, 8 FALSE, 8 UNSURE). 
% distributed equally between both interfaces. 
% The order of the interfaces (interactive / non-interactive) was balanced, and 
The order of presenting the claims during each study session was randomized. 
% The list of claims and corresponding news articles used are shared in the GitHub repository: \url{https://github.com/ixlab-ut/chiir-2022}.

\subsection{Procedure}
\label{sec:procedure}

% ----- fig:experiment_flowchart -----
\begin{figure}[h]
  \centering
  \includegraphics[width=\linewidth]{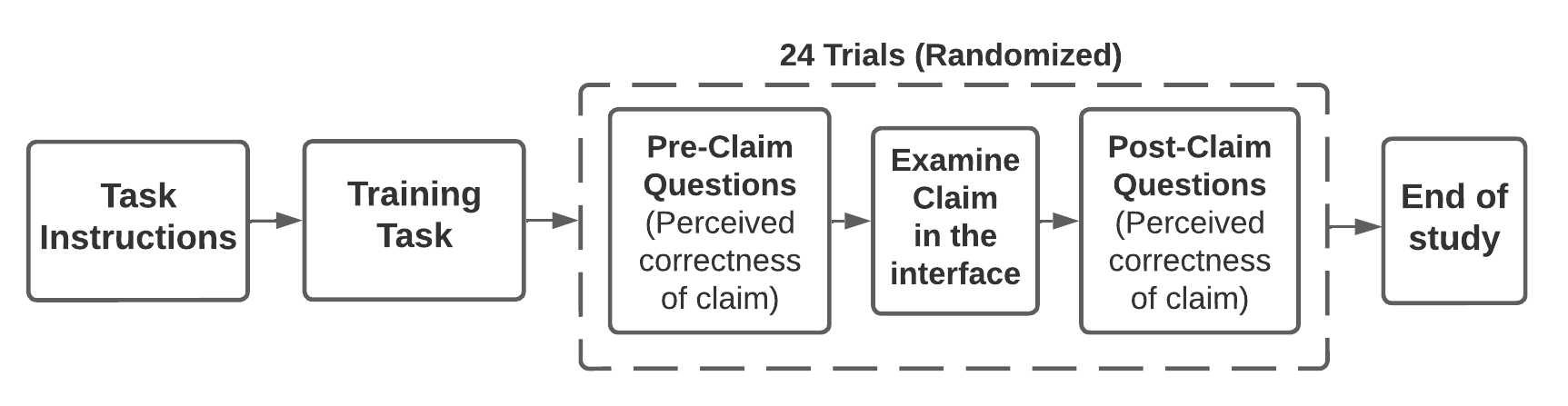}
  \caption{Flowchart of the experimental procedure.}
  \Description{}
  \label{fig:experiment_flowchart}
\end{figure}
% ----- fig:experiment_flowchart -----

The overall procedure of the experimental session is illustrated in Figure ~\ref{fig:experiment_flowchart}. 
Each session started with training task
% (one in interactive, one in non-interactive) 
for participants to get familiar with the interface of the fact-checking system, and the procedure. 
Then the participants started the 24 trials. 
% in one of the two interfaces (experiment blocks), which was randomly chosen and balanced across all participants. 
Each trial consisted of three parts: 
\textit{(i)} Pre-Claim Questions
\textit{(ii)} Examining the claim in the mock fact-checking interface, and 
\textit{(iii)} Post-Claim Questions.

\textbf{Pre-Claim Questions} asked the following:
\begin{itemize}
    % \item Pre-familiarity: \textit{Are you familiar with this claim?: Yes / No / Unsure}
    \item Pre-perceived Correctness: \textit{Do you think the claim is: False / Probably False / Neutral / Probably True / True}
    % \item Pre-confidence: \textit{How confident are you in your assessment of the claim?:Very Low Confidence / Low / Medium / High / Very High Confidence}
    
\end{itemize}

For \textbf{examining the claim}, participants interacted with the interface freely without a time limit.
Participants were also instructed to click on news headlines to open the underlying news articles in a new browser tab, and read it, if they considered it necessary for evaluating the claim. 

\textbf{Post-Claim Questions} asked the following:
\begin{itemize}
    \item Post-perceived Correctness: \textit{After seeing the output of the fact-checking system, do you think the claim is: False / Probably False / Neutral / Probably True / True}
    % \item Post-confidence: \textit{How confident are you in your assessment of the claim?: Very Low Confidence / Low / Medium / High / Very High Confidence}
\end{itemize}

% Before and after viewing each claim in the system, participants indicated their perceived-correctness of the claim in a questionnaire. 
% (which is not analyzed in this short paper). 
% After completing 12 trials in the first interface (block), participants reported their mental workload using the NASA-TLX questionnaire \cite{hart2006nasa}. 
% Then they were allowed to take a five-minute break before resuming the second block (in the other interface).
% NASA-TLX questionnaire was again administered at the end of the block.

\subsection{Measures}
\label{sec:measures}
Our aim was to study cognitive load involved in reading news headlines. 
Previous research \cite{Shi2022effect} found that most fixations on this type of fact-checking interface fell into the headline AOIs. 
This supports the plausibility of studying pupil dilation only on the news headline AOIs. 
So we marked each news headline area (Figure ~\ref{fig:interface}) as an \textit{area of interest} (AOI) for eye-tracking analysis. 
Thus there were five AOIs in the fact-checking interface  (i.e., from the first news headline to the fifth headline).
Javascript function \texttt{Element.getBoundingClientRect()}\footnote{\url{https://developer.mozilla.org/en-US/docs/Web/API/Element/getBoundingClientRect}} was used to get the coordinates for the AOIs. These coordinates were appropriately adjusted to match the coordinates recorded by the eye-tracker.
% All of the AOIs were automatically generated by Javascript. 
% Previous research \cite{shi2022effects}, found that most fixations on this type of fact-checking interface fell into the headline AOIs. 
% This supports plausibility of studying pupil dilation only in the news headline AOIs. 
% All the data were pre-processed by Python and then analyzed in R.
% screenshot under task

\subsubsection{Claim Correctness}
Each claim was selected so as to have a pre-assigned ground-truth correctness value of TRUE, FALSE, or UNSURE (denoted in UPPERCASE). This is the defined as the claim correctness. 
In this research, we wanted to understand the "definitive" behavior on TRUE and FALSE claims first, before trying to tease apart the more complex behavior that may be associated with UNSURE claims. Therefore, 
%the analyses consist of only the trials when users examine the TRUE or FALSE claims.
the analyses in this study only include trials in which users examined TRUE or FALSE claims.

% <FOR_LS> Each claim was selected to have a predetermined ground-truth correctness value of TRUE, FALSE, or UNSURE (denoted in UPPERCASE). This is referred to as claim correctness. We first sought to understand the behavior when evaluating TRUE and FALSE claims, before attempting to analyze the potentially more complex behavior associated with UNSURE claims. As a result, the analyses in this study only include trials in which users examined TRUE or FALSE claims.

\subsubsection{Headline Stance}
For each claim, we collected relevant news articles, which could be supporting or not-supporting the claim. 
Researchers labeled the news Headline Stance based on whether the news article supported the claim or denied the claim, on a 5-item scale: 
-1 (strong deny), 
-0.5 (partially deny), 
0 (neither support nor deny), 
0.5 (partially support), 
1 (strong support).

\subsubsection{Pre- and Post-Perceived Correctness}
Participants' perceived correctness regarding each claim was collected before (Pre-) and after (Post-) they viewed each claim in the fact-checking interface (Section \ref{sec:procedure}).  
Responses to these Pre- and Post-perceived Correctness were on a five-item scale ranging from false to true (denoted in lowercase).

\subsubsection{Evidence correctness}
Evidence correctness denotes the relationship between the headline stance and the claim correctness. 
If the news supports a TRUE claim or denies a FALSE claim, it is categorized as correct evidence. 
In contrast, if the news denies a TRUE claim or supports a FALSE claim, it is categorized as false evidence. 
In this paper, we consider only those news articles that fully supported or fully denied a claim.
\begin{itemize}
    \item \textbf{correct evidence}: headline stance is -1 (strong deny) and claim correctness is FALSE, or, headline stance is 1 (strong support) and claim correctness is TRUE.
    \item \textbf{incorrect evidence}: headline stance is 1 (strong support) and claim correctness is FALSE, or, headline stance is -1 (strong deny) and claim correctness is TRUE.
\end{itemize}

\subsubsection{Headline-Belief-Consistency}

\begin{itemize}
    \item \textbf{headline-belief-consistent}: headline stance is -1 (strong deny) and Pre-Perceived Correctness is false or probably-false; or, headline stance is 1 (strong support) and Pre-Perceived Correctness is true or probably-true.
    \item \textbf{headline-belief-inconsistent}: headline stance is 1 (strong support) and Pre-Perceived Correctness is false or probably false; or, headline stance is -1 (strong deny) and Pre-Perceived Correctness is true or probably true.
\end{itemize}

\subsubsection{Belief Change}
We measured participants' beliefs before and after they checked each claim in the fact-checking system. We grouped their belief change into five categories based on their Pre- and Post-Perceived Correctness, and the Claim Correctness:
\begin{itemize}
    \item \textbf{stay-right}: claim correctness is TRUE and pre-trial and post-trial perceived correctness are both true or probably true; or, claim correctness is FALSE and pre-trial and post-trial perceived correctness are both false or probably false.
    \item \textbf{to-right}: claim correctness is TRUE and post-trial perceived correctness is more towards true than pre-trial; or, claim correctness is FALSE and post-trial perceived correctness is more towards false than pre-trial.
    \item \textbf{stay-neutral}: pre-trial and post-trial perceived correctness are both neutral.
    \item \textbf{to-wrong}: claim correctness is TRUE and post-trial perceived correctness is more towards false than pre-trial; or, claim correctness is FALSE and post-trial perceived correctness is more towards true than pre-trial.
    \item \textbf{stay-wrong}: claim correctness is TRUE and pre-trial and post-trial perceived correctness are both false or probably false; or, claim correctness is FALSE and pre-trial and post-trial perceived correctness are both true or probably true.
\end{itemize}

\subsubsection{Cognitive Load: The Low/High Index of Pupillary Activity (\textit{LHIPA})}
We attempted to apply the LHIPA metric to process pupil dilation data. Pupil dilation during blinks was replaced by 0 according to the eye-tracker detection. After the pre-processing step, we then computed the LHIPA on the raw pupil diameter signal for each headline AOI visits. We observed that the LHIPA value changed significantly when the visit duration was around 1.67s and 6.67s. Because of the high variability associated with the AOI visit duration, the LHIPA metric was not an appropriate indicator of cognitive load, and was not applicable to compare the pupil pupil dilation when processing various news headlines.

% <FOR_LS> In this study, we attempted to use the LHIPA metric to analyze pupil dilation data. Pupil dilation during blinks was replaced with a value of 0 according to the eye-tracker detection. After this preprocessing step, we calculated the LHIPA on the raw pupil diameter signal for each visit to the headline area of interest (AOI). Our results showed that the LHIPA value changed significantly when the visit duration was around 1.67s and 6.67s. However, due to the high variability in AOI visit duration, we found that the LHIPA metric was not a reliable indicator of cognitive load and could not be used to compare pupil dilation when processing different news headlines.

\subsubsection{Cognitive Load: Relative Pupil Dilation (\textit{RPD})}
We calculated pupil dilation based on the raw, high-resolution pupil data recorded at 300Hz. To eliminate individual variability in pupil sizes, we calculated a relative change in pupil diameter from a baseline for each participant. We first excluded the low-quality data (ET-Validity = 4) and the blink data (Blink detected (binary)=1) based on the blink detection algorithm implemented in our eye-tracking software. 
% of iMotions. 
Then we took an average pupil size over all the experimental trials as the pupil diameter baseline (\(P^i_{baseline}\)) and calculated the relative change in pupil diameter (\(RPD^i_t\)) from each pupil measurement (Eq. ~\ref{eqn:rpd}) \cite{gwizdka2017temporal,wang2021pupillary}. We removed data records with diameters that exceeded 
\textpm 3 SDs of the participants' total session average.

\begin{equation}
\label{eqn:rpd}
RPD^i_t = \frac{P_t-P^i_{baseline}}{P^i_{baseline}}
\end{equation}

To calculate the \textit{RPD} for each AOI, we first downsampled the \textit{RPD} to 50Hz using a median filter to minimize the influence of the outliers. Then we excluded the RPD within 0.5 seconds after the interface visit started to reduce the influence of the variability of luminance across the web pages. We assumed that two fixations on a headline did not represent reading and, accordingly, kept the AOI visit which had more than 2 fixations and calculated the \textit{RPD} median for all the AOI visits to a single AOI in each trial.

% -----------------------------------
\section{Results}
\label{sec:results}
\subsection{Testing assumptions}
All assumptions were checked according to the type of statistical testing in this paper. Normality test was conducted before conducting t-test and ANOVA analysis. Bartlett's test was conducted to check for sphericity. %in ANOVA. 
% No assumptions were violated in the analysis.
The results of these tests indicated that no assumptions were violated.

% <FOR_LS> In this study, all assumptions were carefully checked prior to statistical testing. Normality tests were conducted for t-tests and ANOVA, and Bartlett's test was used to check for sphericity in ANOVA. The results of these checks indicated that no assumptions were violated in the analysis.

\subsection{AOI position}
\label{sec:AOI_position}

% ----- Figure 3 -----
\begin{figure}[h]
  \centering
  \includegraphics[width=\linewidth]{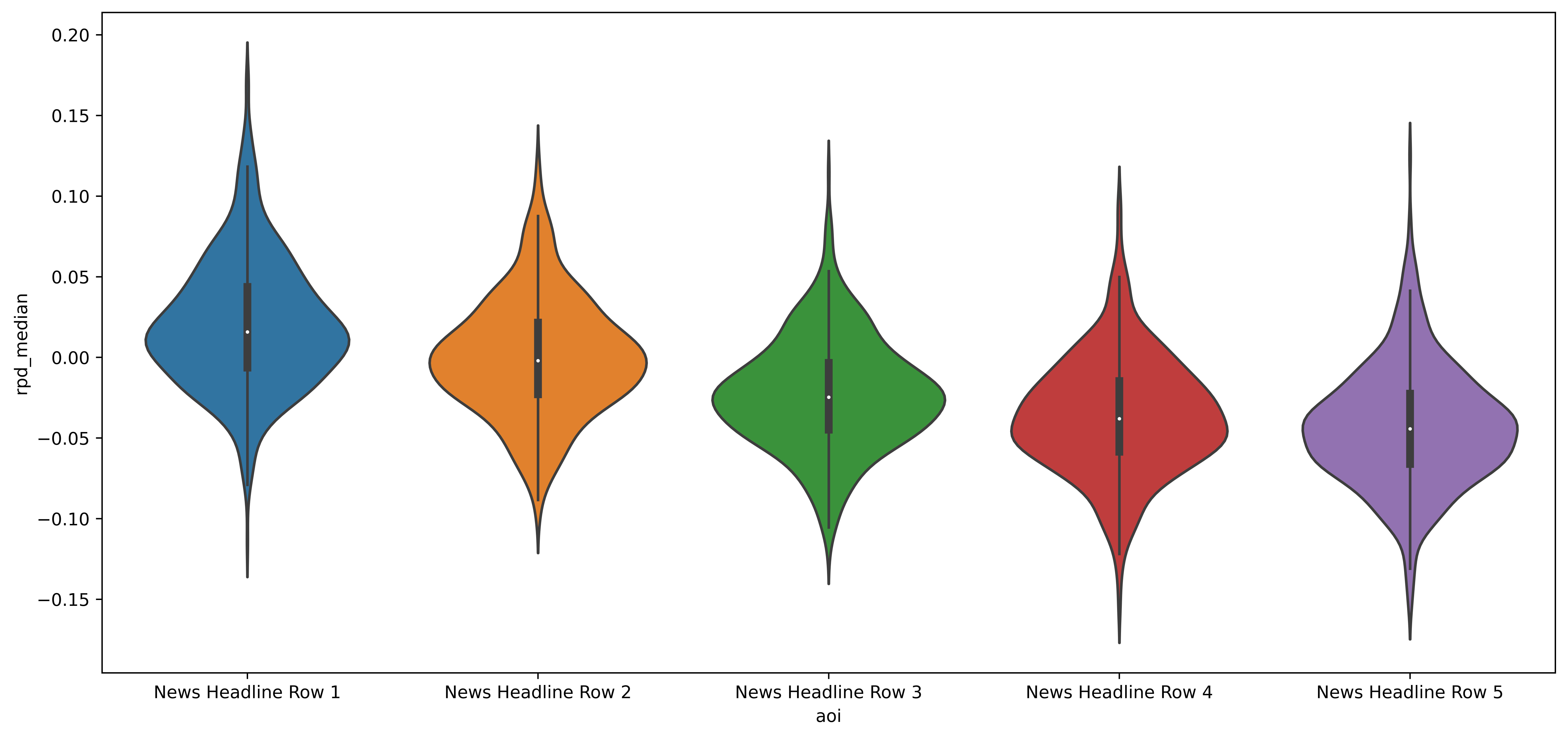}
  \caption{Relative pupil dilation (\textit{RPD}) in the areas of interest (AOIs). From the left to the right it represents the AOIs of the news headlines that are from the first row to the fifth row of the fact-checking interface.}
  % \Description{Violin graph for \textit{RPD} in different AOIs. \textit{RPD} is largest in the first row, then it decreases in the following rows.}
  \label{fig:aoi_position}
\end{figure}
% ----- Figure 3 -----

Figure ~\ref{fig:aoi_position} shows that relative pupil dilation (\textit{RPD}) was largest when participants were reading the news headlines in the first row. Then the \textit{RPD} decreased as they read the headlines in the following rows. A one-way ANOVA showed that the effect of headline position (i.e. the rank of headline in the interface) was significant, $F(4,3004)=321.9, p<.05$. A post hoc Tukey's HSD test showed that all groups differed significantly at $p<.05$.
% combine with the violin chart, the rpd decreases from top to the bottom

\subsection{Claim correctness and headline stance}
\label{sec:correct_stance}

% ----- Figure 4 -----
\begin{figure}[h]
  \centering
  \includegraphics[width=\linewidth]{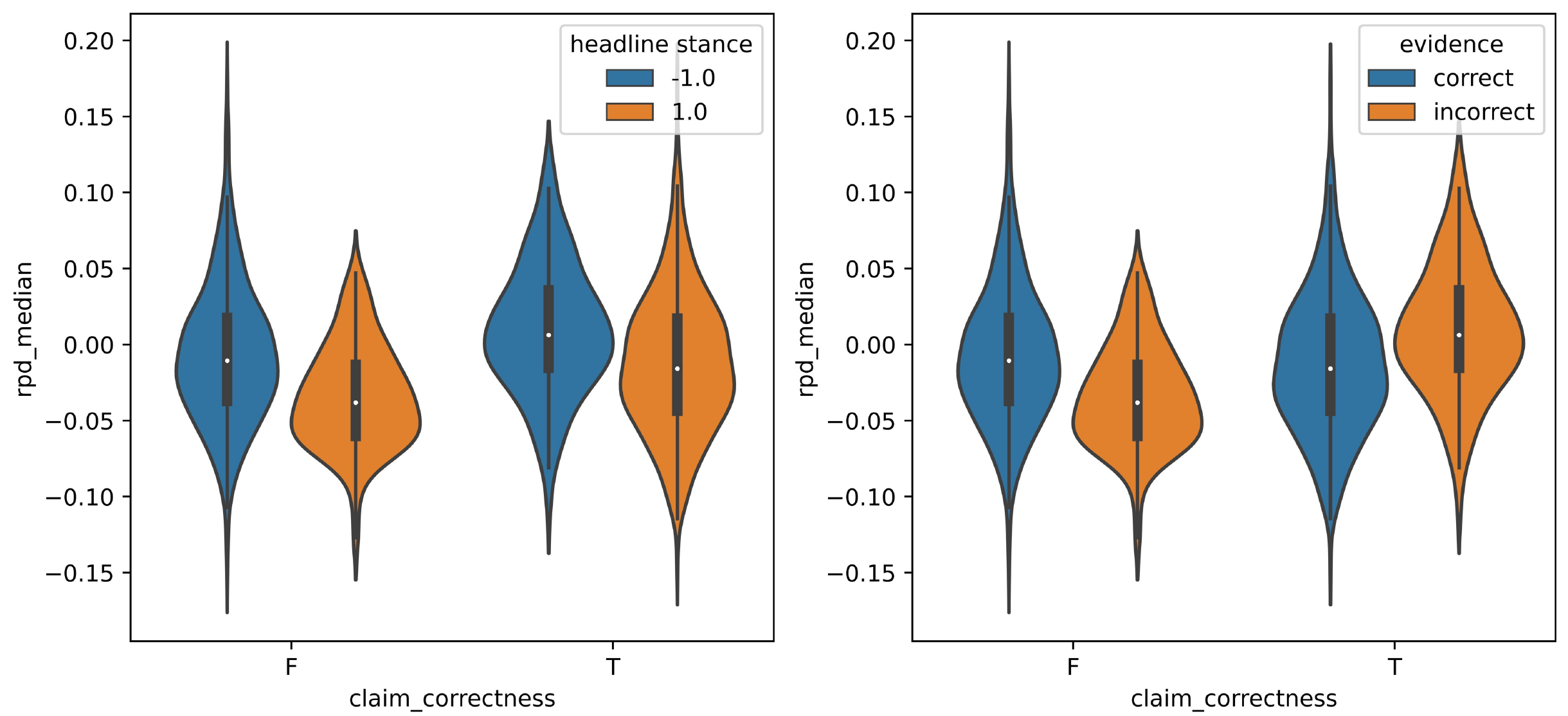}
  \caption{(a) Distribution of \textit{RPD} of the news headline AOIs as a function of headline stance (-1: headline denies the claim; 1: headline supports the claim) and claim correctness (TRUE or FALSE claim). (b) Distribution of \textit{RPD} of the news headline AOIs as a function of the evidence correctness (correct or incorrect) and claim correctness (TRUE or FALSE claim).}
  % \Description{Violin charts for \textit{RPD} in news headline AOIs. In figure(a), \textit{RPD} is larger in both TRUE and FALSE claims when the headline denys the claim. In figure(b), \textit{RPD} is larger in TRUE claims when the headline stance is inconsistent with the claim correctness, while in FALSE claim, relative pupil dilation was larger in FALSE claims when the headline stance is consistent with the claim correctness.}
  \label{fig:claim_hdl_consist}
\end{figure}
% ----- Figure 4 -----

Figure ~\ref{fig:claim_hdl_consist}(a) indicates that larger \textit{RPD} was on the news headline AOIs that denied the claims. \textit{RPD} was generally larger when checking TRUE claims compared to FALSE claims. A two-way ANOVA was conducted to examine the effects of headline stance and claim correctness on \textit{RPD}. Both claim correctness, F(1,34)=15.54, p<.05, and headline stance, F(1,34)=31.98, p<.05, had significant main effects on \textit{RPD}. However, the interaction effects were not significant. 
A post hoc Tukey's HSD test showed that all groups differed significantly at p<.05.
The \textit{RPD} was larger when participants were reading the news headlines that denied the claim, in both TRUE and FALSE claim group.
The \textit{RPD} was larger when participants were checking TRUE claims, no matter if they were reading news headlines denying or supporting a claim.

Figure ~\ref{fig:claim_hdl_consist}(b) illustrates that \textit{RPD} was larger when participants were checking the incorrect evidence for TRUE claims, and when checking the correct evidence for FALSE claims.
\textit{RPD} was tested by a two-way ANOVA with two levels of claim correctness (TRUE, FALSE) and two levels of evidence correctness (correct, incorrect). The main effect of the claim correctness was significant, F(1,34)=15.54, p<.05. The main effect of the evidence correctness was not significant. However, the interaction of claim correctness and evidence correctness was significant, F(1,34)=31.98, p<.05. 
A post hoc Tukey's HSD test showed that all the groups differed significantly at p<.05. 
When participants were checking TRUE claims, the \textit{RPD} was larger in the incorrect evidence group compared to the correct evidence group, while checking FALSE claims, the \textit{RPD} was lower in the incorrect evidence group compared to the correct evidence group.
When participants were reading correct evidence, the \textit{RPD} was larger in the FALSE claim group compared to the TRUE claim group, while when participants were reading incorrect evidence, the \textit{RPD} was larger in the TRUE claim group compared to the FALSE claim group.

\subsection{Prior belief and headline stance}
\label{sec:belief_stance}

% ----- Figure 5 -----
\begin{figure}[h]
  \centering
  \includegraphics[width=\linewidth]{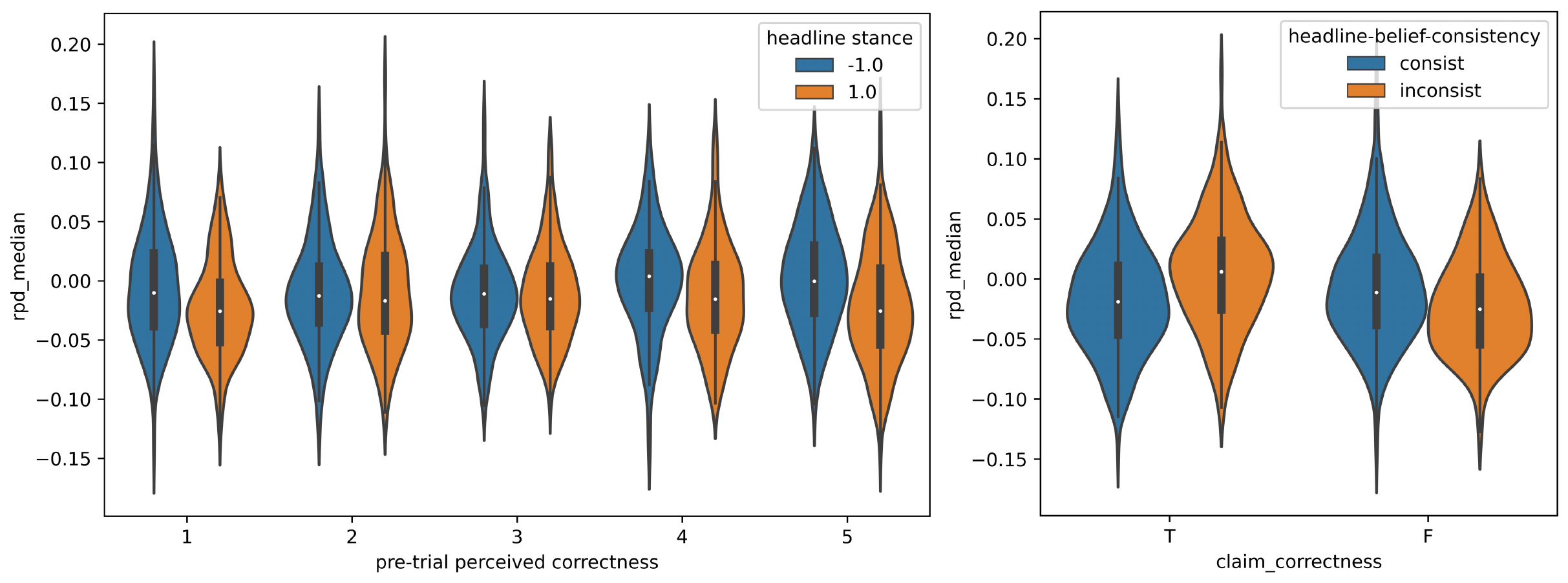}
  \caption{(a) Distribution of \textit{RPD} on the news headline AOIs as a function of the perceived correctness before examining the claim in the system (1 to 5: False to True) and the headline stance (-1: headline denies the claim; 1: headline supports the claim). (b) Distribution of \textit{RPD} of the news headline AOIs as a function of the headline-belief-consistency (consistent or inconsistent) and claim correctness (TRUE or FALSE claim).}
  % \Description{Violin graphs for \textit{RPD} in news headline AOIs. In figure(a), \textit{RPD} is larger when they are looking at the denying news headlines when the prior belief is totally true or totally flase. In figure (b), \textit{RPD} is larger in TRUE claims when the headline stance is consistent with the prior belief, while \textit{RPD} is larger for TRUE claims when the headline stance is inconsistent with prior beliefs.}
  \label{fig:belief_hdl_consist}
\end{figure}
% ----- Figure 5 -----

Figure ~\ref{fig:belief_hdl_consist}(a) shows that the largest difference in \textit{RPD} between the news headlines that supported and those that denied the claim was when participants' prior belief was true or false. That difference was smaller when their prior belief was neutral. Therefore we looked further into the relationship between headline stance and the prior belief in two directions (i.e., the perceived correctness of the claim was either true or false). Figure ~\ref{fig:belief_hdl_consist}(b) shows that the \textit{RPD} was higher when headline-belief was inconsistent in TRUE claim groups and when headline-belief was consistent in FALSE claim groups.
A two-way ANOVA was conducted to examine the effects of the headline-belief consistency and the claim correctness on \textit{RPD}. The claim correctness had a significant main effect, F(1,35)=8.42, p<.05, while the headline-belief consistency had no significant effect. The interaction effects of claim correctness and headline-belief consistency were significant, F(1,35)=23.31, p<.05.
A post hoc Tukey's HSD test showed that all the groups differed significantly at p<.05.
When checking TRUE claims, the \textit{RPD} was larger in the headline-belief inconsistent group compared to the headline-belief consistent group, while when checking FALSE claims, the \textit{RPD} was lower in the headline-belief inconsistent group compared to the headline-belief consistent group.
When headline-belief was consistent, the \textit{RPD} was larger in the FALSE claim group compared to the TRUE claim group, while when headline-belief was inconsistent, the \textit{RPD} was larger in the TRUE claim group compared to the FALSE claim group.

\subsection{Belief change}
\label{sec:belief_change}

% ----- Figure 6 -----
\begin{figure}[h]
  \centering
  \includegraphics[width=\linewidth]{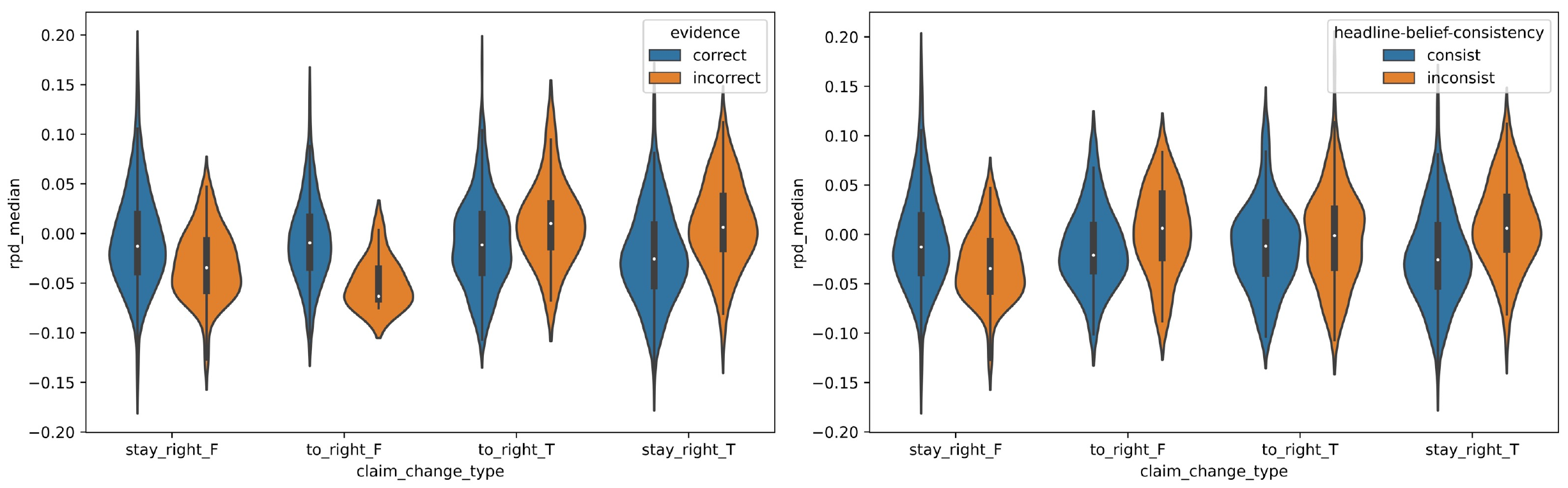}
  \caption{(a) Distribution of \textit{RPD} of the news headline AOIs as a function of the evidence correctness (correct or incorrect) and the belief change in both claims (stay/to-right in TRUE/FALSE claims). (b) Distribution of \textit{RPD} of the news headline AOIs as a function of belief-headline-consistency (consistent or inconsistent) and the belief change in both claims (stay/to-right in TRUE/FALSE claims).}
  
  % \Description{The violin graphs for \textit{RPD} in news headline AOIs. In the figure (a), the \textit{RPD} of the participants who were checking FALSE claims, no matter their belief stay-right or change to-right, are larger when the news is good evidence. On the contrast, the \textit{RPD} of the participants who were checking TRUE claims, no matter their belief stay-right or change to-right, are larger when the news is bad evidence. In figure (b), \textit{RPD} has apparently difference when the users' believe stay-right, no matter checking TRUE or FALSE claim.}
  \label{fig:change_trend}
\end{figure}
% ----- Figure 6 -----

% Figure ~\ref{fig:change_trend} demonstrates that the central tendency of \textit{RPD} didn't differ a lot but the distribution of \textit{RPD} when the participants' belief stayed to be wrong or moved to the wrong direction is different from the distribution of other belief change conditions. 
A one-way ANOVA $(F(4,2175)=0.61)$ indicated that the \textit{RPD} was not significantly different between belief change conditions. Therefore, the change of the belief did not significantly influence the \textit{RPD}. In our lab experiment, participants maintained their correct beliefs (stay-right) in 44.22\% of the trials, and corrected their beliefs (to-right) in 46.56\% of the trials. Only in 9.22\% of the trials, participants stayed neutral, or remained incorrect (stay-wrong), or changed their beliefs to incorrect (to-wrong).

% here to put the t-test and mann whitney u test result
Furthermore, we checked the impact of the evidence correctness and the headline-belief-consistency on the \textit{RPD} of the participants whose beliefs were corrected (to-right) or remained correct (stay-right). 
Figure ~\ref{fig:change_trend}(a) shows that \textit{RPD} differed between reading the headlines that are correct evidence and incorrect evidence within each belief change and claim correctness combination group (i.e., to-right in TRUE claims, to-right in FALSE claims, stay-right in TRUE claims, stay-right in FALSE claims).
\textit{RPD} when reading correct evidence had small differences between belief change and claim correctness combination groups, while \textit{RPD} when reading incorrect evidence had larger differences between belief change and claim correctness combination groups.
when reading the incorrect evidence, the \textit{RPD} differed more between change trend groups. 
A paired-sample t-test was conducted to compare \textit{RPD} correct evidence and incorrect evidence conditions within each belief change and claim correctness combination group. There were significant differences in \textit{RPD} between reading correct and incorrect evidence in all the belief change groups: stay-right for FALSE claims, $t(28)=4.44, p<.05$, to-right for FALSE claims, $t(8)=3.97, p<.05$, to-right for TRUE claims, $t(13)=-2.36, p<.05$, stay-right for TRUE claims, $t(29)=-5.25, p<.05$. 

% _____________ tab_pdf_mannwhitneyu _____________
\begin{table}[!thbp]
% \captionsetup[subfloat]%{farskip=0pt,captionskip=0pt}
\centering
\caption{
Mann-Whitney U tests to determine whether there were significant differences between maintaining correct beliefs (stay-right) vs. correcting beliefs (to-right) (Section \ref{sec:belief_change}),
\textbf{(a)} for different combinations of Claim-Correctness and Evidence Correctness, and
\textbf{(b)} for different combinations of Claim-Correctness and Headline-belief Consistency.
(*$p<.05$, **$p<.01$, ***$p<.001$)
}
\label{mannwhitneyu}
\includegraphics[clip, trim=0 0 0 0, width=\linewidth]{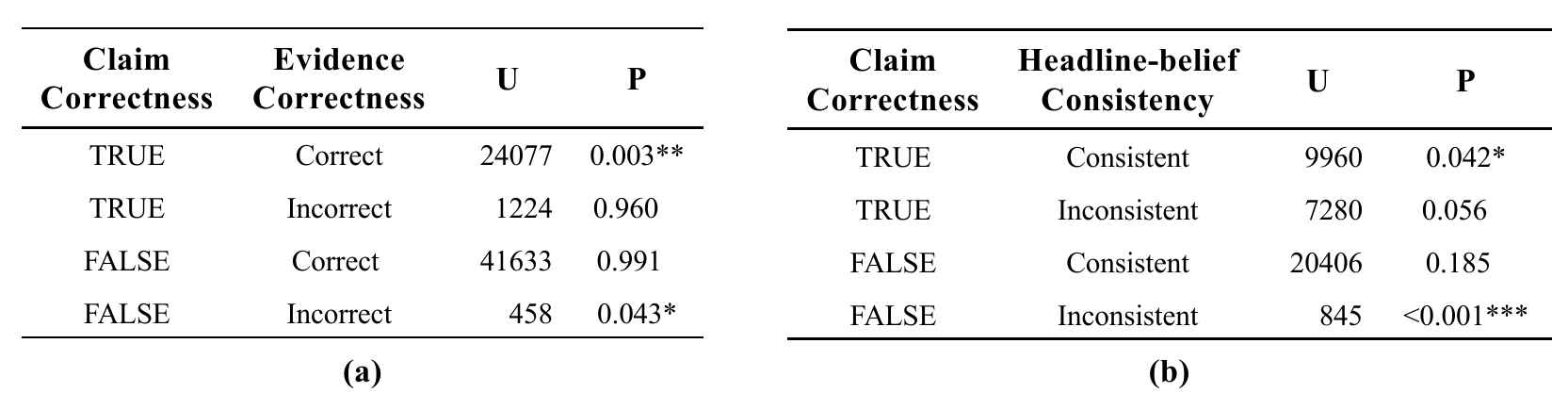} %trim - left, bottom, right, top

\end{table}
% _____________ tab_pdf_mannwhitneyu _____________

% A Mann-Whitney U test was conducted to determine whether there was a difference when the participants' belief stayed right and  moved to-right in each evidence and claim groups. When the participants checked the correct evidence in FALSE claims, \textit{RPD} didn't show significant difference between two belief change conditions. When they checked the incorrect evidence in FALSE claims, \textit{RPD} is significantly greater (U= 458.0,p<.05) when the belief stayed in right comparing to changing to-right. When the participants checked the correct evidence in TRUE claims, \textit{RPD} is significantly greater (U=24077.0, p<.05) when the belief changed to-right comparing to staying in the right. When the participants checked the incorrect evidence in TRUE claims, \textit{RPD} didn't show significant difference between two belief change conditions.
Figure ~\ref{fig:change_trend}(b) shows that the differences between the headline-belief consistent group and inconsistent group.
\textit{RPD} in headline-belief consistent group were larger than headline-belief inconsistent group when participants maintained their correct beliefs (stay-right) in FALSE claims.
\textit{RPD} in headline-belief inconsistent group were larger than headline-belief consistent group when participants changed to the correct belief (to-right) in both TRUE and FALSE claims, and when participants maintained their correct beliefs (stay-right) in TRUE claims.
A paired-sample t-test was conducted to compare \textit{RPD} in headline-belief consistent and inconsistent groups within each belief change and claim correctness combination group. There were significant differences in \textit{RPD} in two of the belief change groups: stay-right for FALSE claims, $t(28)=4.44, p<.05$, to-right in TRUE claims, $t(13)=-5,25, p<.05$.
A Mann-Whitney U test was conducted to determine whether there was a difference when the participants' beliefs stayed right and  moved to the right in each evidence correctness or headline-belief-consistency groups and claim groups. Table ~\ref{mannwhitneyu} demonstrates the results. The Mann-Whitney U test and paired-samples t-test indicated that \textit{RPD} was significantly larger when the participants corrected their beliefs (to-right) especially in the headline-belief consistent group for TRUE claims, and in the headline-belief inconsistent group for FALSE claims.

% -----------------------------------
\section{Discussion}
\label{sec:discussion}

In this study, we investigated how the cognitive load is impacted in the fact-checking context and if it is related to users' belief change. We conducted a within-subject, lab-based, quasi-experiment, in which we manipulated the evidence correctness (correct evidence, incorrect evidence), the headline-belief-consistency (consistent, inconsistent), and measured participants' belief change (stay-right, stay-neutral, stay-wrong, to-right, to-wrong). We evaluated the cognitive load when participants read the news headlines by measuring pupil dilation on the headline AOIs.

We found that \textit{RPD} is the highest when users read the news headlines in the top row, and that the \textit{RPD} decreases on the lower positioned headline rows. This suggested that the cognitive load is higher when people are processing top headlines. This could be because of the position bias \cite{azzopardi2021cognitive}, where highly ranked results tend to attract more attention, reading and clicks. Since the news headlines with different stances were randomly assigned to the headline row positions, the position bias does not influence our hypothesis testing of other factors potentially impacting cognitive load.

\textit{Evidence correctness.} 
We found that higher cognitive load is imposed when reading news headlines that are denying the claim. That is, higher cognitive load is required when users read incorrect evidence for TRUE claims and correct evidence for FALSE claims. This finding supports H1 when users are fact-checking a TRUE claim but provides no support for H1 when users are fact-checking a FALSE claim.
Additionally, when users are checking TRUE claims, higher cognitive load is imposed regardless of the evidence correctness.

\textit{Headline-belief-consistency.}
When checking the relationship between headline stance and prior beliefs, the results indicate that cognitive load differs more between supportive and unsupportive headlines when the users perceived the claim correctness as true or false, instead of being neutral. The observation could be explained by that users would like to check all the information and a similar amount of cognitive load is imposed to read both supportive and unsupportive news since they do not have a prior tendency to claim correctness.
Furthermore, we looked into the association between cognitive load and the news headlines when the users held non-neutral prior beliefs. We found evidence for the effect of headline-belief consistency on cognitive load. Reading headline-belief inconsistent news imposed higher cognitive load when checking TRUE claims, while reading headline-belief consistent news imposed higher cognitive load when checking FALSE claims. This finding supports H2 for TRUE claims but denies H2 for FALSE claims.

\textit{Belief change.}
We did not find a relationship between cognitive load and belief change, which denies H3. This result is not aligned with previous findings on discerning misinformation performance \cite{Mirhoseini_Early_Hassanein_2022}. It is possible that our findings are due to the participants using the fact-checking system in the experiment and being more aware of the tasks to discern misinformation, hence they invest similar mental effort to process all news headlines shown in the interface and to judge the claim correctness. 
Additionally, the belief change results suggest that participants generally performed well on the fact-checking tasks. Among all the experiment trials, users' kept their correct beliefs (stay-right) or corrected them (to-right) in more than 90\% trials. This indicates that the fact-checking system helped users to discern misinformation.
In the analysis within the stay-right and to-right belief change groups, the results cross-validated the findings in evidence correctness and headline-belief-consistency - H1 and H2 are supported for TRUE claims and rejected for FALSE claims. Moreover, the cognitive load was higher when users' corrected their beliefs (to-right) compared to when they maintained their beliefs (stayed-right) when users were reading headline-belief-consistent news for TRUE claims and headline-belief-inconsistent for FALSE claims. This implies that checking news headlines when users' beliefs were corrected (to-right) imposed higher cognitive load than checking news headlines when they maintained their correct beliefs (stay-right).

In summary, when users were reading news headlines for TRUE claims, our proposed hypotheses H1 and H2 were supported, while when users were reading news headlines for FALSE news, H1 and H2 were not supported. There was not enough evidence in our study to support H3. We found that incorrect evidence and headline-belief inconsistency may not always impose higher cognitive load. Instead, the cognitive load level imposed by reading headlines appeared to be associated with the claim correctness. When checking TRUE claims, higher cognitive load was imposed when users read incorrect evidence or read headline-belief-inconsistent news. When checking FALSE claims, higher cognitive load was imposed when users read correct evidence or read headline-belief-consistent news. The findings plausibly indicated that people tended to engage more with the news they believed in when they were checking FALSE claims, while they engaged more with the news that countered their belief when they were checking TRUE claims. We also found that cognitive load did not significantly differ between belief change conditions, which suggests that the fact-checking tasks imposed similar level of cognitive load regardless how people's beliefs changed. However, the results indicate that a higher cognitive load was imposed when users corrected their beliefs and when users were reading headline-belief-consistent news for TRUE claims, or headline-belief-inconsistent for FALSE claims.

This research develops an understanding of cognitive load in discerning misinformation in realistic scenarios. Based on previous research on the association between cognitive load and reading and identifying misinformation \cite{Mirhoseini_Early_Hassanein_2022}, we studied how people process misinformation when they encounter several headlines (or news search results) at the same time while examining a single claim. Our findings suggest that cognitive load is imposed differently when checking true claims versus false claims. 
Previous research suggests that information system should encourage people to engage more cognitive effort (System 2), which could help them to identify misinformation more effectively \cite{rieger2021item}. Meanwhile, we need to prevent the cognition to be overloaded which could drive the users back to utilize their heuristic \cite{whelan2020applying}. 
Our study implies that there are nuances in cognitive load when people are processing information with different claim correctness, evidence correctness, and headline-belief-consistency.
As suggested in \cite{littrell2022not}, different kinds of misinformation could invoke different information behavior. In practical system design, we should not simply increase or decrease the cognitive load, but instead seek to calibrate the cognitive load with respect to the information context. We need to adopt a more nuanced approach to nudge people to discern misinformation, such as providing personalized labels or explanations to remind people to pay attention to the misinformation at the appropriate condition.

% <FOR_LS> This research aims to improve our understanding of cognitive load in the context of discerning misinformation in realistic situations. Based on previous studies that have explored the relationship between cognitive load and reading and identifying misinformation \cite{Mirhoseini_Early_Hassanein_2022}, we examined how people process misinformation when confronted with multiple headlines or news search results that contain a single claim. Our findings suggest that cognitive load is imposed differently when evaluating true and false claims. Previous research has suggested that information systems should encourage people to engage more cognitive effort (System 2) in order to identify misinformation more effectively \cite{rieger2021item}. However, it is also important to avoid overloading cognition, which can lead people to rely on heuristics \cite{whelan2020applying}. Our study highlights the complexities of cognitive load when people are processing information with different levels of claim correctness, evidence correctness, and headline-belief consistency. As noted in \cite{littrell2022not}, different types of misinformation can evoke different information behaviors. In practical system design, it is important to consider the specific context of the information and adopt a nuanced approach to nudging people to discern misinformation, such as providing personalized labels or explanations to remind people to pay attention to potential misinformation under the appropriate conditions.

Our study has some limitations. We only observed the effect of evidence correctness, headline-stance-consistency, and belief change on the cognitive load when reading the news headlines. The cognitive load could also be impacted by users' familiarity and the knowledge level of the claim topic. Even though we have excluded the participants with expert topic familiarity of the content based on self-reported information, there's a possibility that people are not aware of their expertise in the topic. Since higher familiarity could impose lower the cognitive load \cite{jen2017examining}, this limitation would impact the internal validity of our research.
Additionally, the claims and news headlines were pre-selected to conduct the controlled within-subject experiment. Future work should include using sets of claims on different topics and investigating cognitive load in the context of naturally generated fact-checking tasks. Another limitation of this study is that we only measured pupil dilation when they were looking at the news headline AOIs. It would therefore be interesting to measure pupil dilation when they read the full news articles and compare the cognitive load variations between distinct news conditions. Lastly, 
the eye-tracking sequences in the experiment are relatively short. This renders them inapplicable to use the LHIPA technique to process and analyze the pupil dilation. Future work could improve the experimental design and allow for other pupillary response measurements (i.e., LHIPA) to reflect cognitive load with higher accuracy \cite{duchowski2020low}, or even other physiological measures, such as Electroencephalography (EEG) \cite{antonenko2010using}.

% % _____________ tab_pdf_example _____________
% \begin{table}[!thbp]
% % \captionsetup[subfloat]%{farskip=0pt,captionskip=0pt}
% \centering
% \caption{
% Deep learning (of the human kind)
% versus traditional (also often online) classroom  practices. 
% Compiled from \cite{sawyer2005cambridge, cope2017elearningc}.
% }
% \label{tab_pdf_example}
% \includegraphics[clip, trim=0 0 0 0, width=\linewidth]{figs/tab_pdf_example.pdf} %trim - left, bottom, right, top

% \end{table}
% % _____________ tab_pdf_example _____________

% -----------------------------------
\section{Conclusion}
\label{sec:conclusion}
We presented results from a within-subject, lab-based, quasi-experiment with eye-tracking in which we examined how cognitive load is impacted by reading news headlines in a fact-checking context (i.e., by the evidence correctness and users' prior beliefs), and how it is related to people's belief change and their misinformation judgment.
We found that incorrect evidence and headline-belief inconsistency imposed higher cognitive load when people were checking true claims, while correct evidence and headline-belief consistency imposed higher cognitive load when people were checking false claims. Additionally, cognitive load was not significantly different when people's beliefs changed. 
By developing the understanding of the cognition in discerning misinformation in a realistic scenario, the findings contribute to designing future information systems that support curbing of misinformation spread via appropriate technical and cognitive interventions.

\begin{acks}
This research was completed under UT Austin IRB study 2017070049 and supported in part by Wipro, the Micron Foundation, the Knight Foundation, and by Good Systems\footnote{http://goodsystems.utexas.edu/} , a UT Austin Grand Challenge to develop responsible AI technologies. The statements made herein are solely the opinions of the authors and do not reflect the views of the sponsoring agencies.
\end{acks}

\bibliographystyle{ACM-Reference-Format}
\bibliography{references}

\end{document}